\documentclass[twocolumn, groupedaddress]{revtex4-1}

\usepackage{graphicx} 

\begin{document}

\title{Characterising equation of state and optical properties \\of dynamically pre-compressed materials}

\author{M. Guarguaglini}
\email[Electronic mail: ]{marco.guarguaglini@polytechnique.edu}
\author{J.-A. Hernandez}
\author{A. Benuzzi-Mounaix}
\author{R. Bolis}
\author{E. Brambrink}
\author{T. Vinci}
\author{A. Ravasio}
\email[Electronic mail: ]{alessandra.ravasio@polytechnique.edu}

\affiliation{LULI - \'Ecole Polytechnique, CNRS, CEA, Universit\'e Paris-Saclay, route de Saclay, 91128 Palaiseau cedex, France}
\affiliation{Sorbonne Universit\'e, Laboratoire d'Utilisation des Lasers Intenses, CNRS, Campus Pierre et Marie Curie, 4 place Jussieu, 75252 Paris cedex 05, France}

\date{\today}

\begin{abstract} 
Characterising materials at pressures of several megabar and temperatures of a few thousand Kelvin is critical for the understanding of the Warm Dense Matter regime and to improve planetary models as these conditions are typical of planets' interiors.
Today, laser-driven shock compression is the only technique to achieve multimegabar pressures, but the associated temperatures are too high to be representative of planetary states. Double-shock compression represents an alternative to explore lower temperatures.
Here we present a method to create well-controlled double-shocked states and measure their thermodynamic state and optical reflectivity using standard optical diagnostics (Doppler velocimetry and optical pyrometry) in a laser-driven shock experiment. This method, which does not require the support of hydrodynamical simulations, is based on the application of generalised Rankine-Hugoniot relations together with a self impedance mismatch technique.
A validation experiment has been performed at the LULI2000 facility (\'Ecole Polytechnique, France) on a sample of $\alpha$-quartz. A temperature $60 \%$ lower than along the principal Hugoniot has been obtained at 7.5 Mbar.
\end{abstract}




\maketitle


\section{Introduction}
\label{sec:introduction}

The characterisation of the physical properties of materials at pressures of several megabar (1 Mbar = 100 GPa) and temperatures of a few thousand Kelvin is critical for our understanding of the Warm Dense Matter regime \cite{koenig05} and has applications in many branches of physics, especially planetary science \cite{saumon04, stixrude08, guillot99}. Indeed, models describing birth, evolution, and structure of planets and the generation of their magnetic fields require the precise characterisation of equation of state, phase diagram, and electrical conductivity of some key components at planetary interiors conditions.

Exploring a wide range of pressure and temperature conditions requires the use of different compression techniques. The most effective method to reach a 1 - 10 Mbar pressure range is laser-driven shock compression \cite{koenig95}. In this context, a first approach consists in generating a single shock wave and let it propagate through the sample, initially at ambient conditions. This allows exploring principal Hugoniot states \cite{forbes} at very high temperatures ($10^4 - 10^5$ K) and limited compression factors (below 10 \cite{zeldovich}). This regime is mostly characterised by fluid, conducting, dense-plasma states \cite{laudernet04, clerouin05, french10, mcwilliams12}.

At lower temperatures ($10^3 - 10^4$ K), materials exhibit rich phase diagrams and chemistry-driven phenomenon such as molecular dissociation and phase separation \cite{chau11}. Their characterisation is critical for planetary modelling since planetary interior profiles are mostly adiabatic, with temperatures of some thousand Kelvin \cite{guillot05}. In this regime, calculations predict an exotic superionic phase for water \cite{cavazzoni99, goldman05, french09, hernandez16, hernandez18}, ammonia \cite{cavazzoni99, bethkenhagen13}, and water-ammonia mixtures \cite{bethkenhagen15}. In this peculiar state, the high proton diffusion produces liquid-like electrical conductivities, with possible consequences on the magnetic field expression in icy giant planets. An experimental evidence of superionic water has been only recently found \cite{millot18}. 
Another important planetary application in this low-temperature regime is the determination of the melting lines and polymorphic transitions of silicates, which are critical to infer the evolution history and internal structure of terrestrial planets such as super-Earths \cite{umemoto06, millot15}. 

Exploring this low-temperature, multimegabar-pressure regime with a shock-compression approach requires the initial density of the sample to be increased. Indeed, when a shock propagates in a denser state it transfers a larger amount of its energy to compress it instead of heating \cite{zeldovich}. 

A first way to modify the initial density is to consider, if they exist, the different stable or metastable solid polymorphs at ambient conditions \cite{millot15}. Second, the sample may be pre-compressed statically (e.g. in a diamond anvil cell) and maintained at high pressure prior to the shock \cite{jeanloz07}. Third, the initial density can be increased dynamically based on several approaches: \textit{(i)} confining the sample with a higher-density material so to obtain a shock reverberation \cite{nellis99, boehly04, knudson12}; \textit{(ii)} shaping the laser pulse to compress the sample along a quasi-isentropic path \cite{swift05, smith14}; \textit{(iii)} using several laser pulses to produce multiple shocks \cite{benuzzi04}. 

Pre-compression methods must at the same time allow to reach the desired conditions and to precisely characterise them. Static pre-compressions with diamond anvil cells make very precise measurements of the reached state possible but are currently limited to $12$ GPa when coupled with laser shock compression \cite{brygoo15, brygoo17}. Moreover, most of time a substantial fraction of the laser energy is spent to sustain the shock through the thick anvil. Reverberating shock compression reaches much lower temperatures than along the principal Hugoniot, although hydrodynamical simulations (necessarily based on an equation of state model) often have to be used to infer the reached conditions \cite{denoeud16}, but model-independent results can be obtained \cite{boehly04}. Employing simulations is generally necessary if quasi-isentropic paths are followed \cite{xue18}, even if absolute stress-density relations can be measured \cite{smith14}.

In this paper we present an alternative approach based on double-shock compression in which the thermodynamic state and optical reflectivity of re-shocked states are precisely characterised using standard optical diagnostics, such as Doppler velocimeters (VISARs) and an optical pyrometer (SOP). This technique is relevant for transparent materials, such as silica, water, and liquid ammonia. The measurement of the thermodynamic state is based on a generalisation of the Rankine-Hugoniot relations \cite{birnboim10} and on a self impedance mismatch analysis. Reflectivity is measured exploiting the VISAR data.

Our technique requires the knowledge of the principal Hugoniot and of the refractive index of the material in the transparent portion of the Hugoniot. These are well known for many materials of planetary interest \cite{sano11, knudson04, kimura15, zha94, batani15}. This requirement does not limit the application field of this technique since these quantities can be measured in a preliminary part of the experimental campaign. The use of simulations is not required to infer the reached state.

The validity of the employed technique has been tested at the LULI2000 laser facility (\'Ecole Polytechnique, France). A pressure of 7.5 Mbar was achieved in shocked $\alpha$-quartz, initially brought to 0.33 Mbar (33 GPa) by a first weak compression. The achieved temperature was 12000 K, $60\%$ lower than expected along the Hugoniot at the same pressure. This technique extends the possibilities to investigate these high pressure - moderate temperature regimes to a wide class of medium-size laser facilities.

This paper is organised as follows. In Section \ref{sec:technique_principles} we will first outline the principles of the double-shock compression technique by describing the hydrodynamic path and explaining the self impedance mismatch technique. In Section \ref{sec:simulations} we will then present the preliminary study done with hydrodynamical simulations to design targets and laser pulses. Section \ref{sec:exp_application} will describe the experimental setup of the LULI2000 campaign and the analysis of diagnostics outputs to extract velocities, reflectivities, and temperatures. Section \ref{sec:results} and \ref{sec:conclusions} are respectively dedicated to the statement and discussion of the results and our conclusions.

\begin{figure*}
\includegraphics[width=0.8\textwidth]{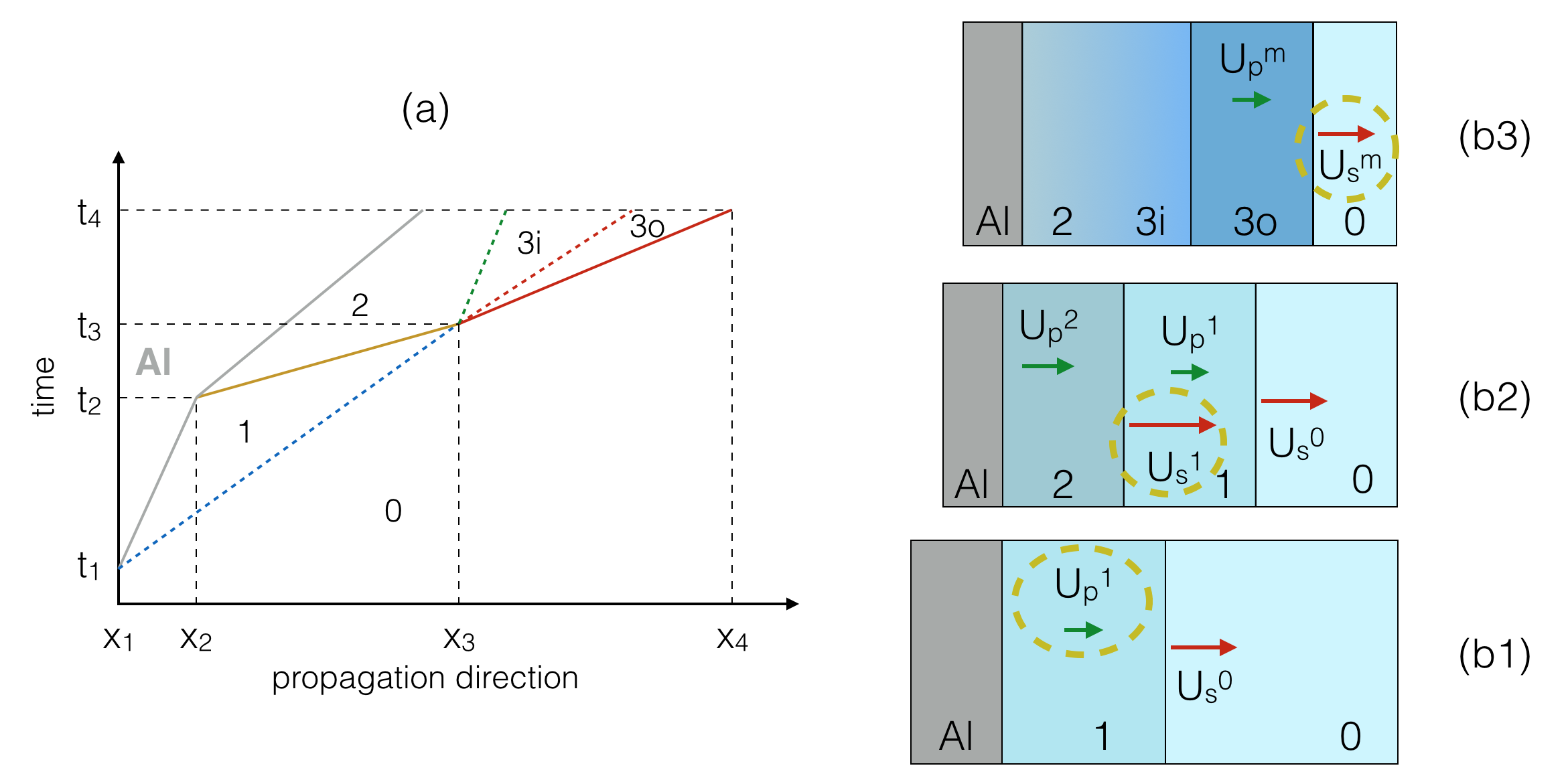}
\caption{
Drive lasers come from the left, diagnostics are on the right side. 
\textbf{(a).} Shock propagation history inside the sample. $x_1$ and $x_4$ are the initial position of the aluminum/sample interface and of the rear side of the sample, respectively. The grey line is the path of the aluminum/sample interface. The blue dashed line is the first shock front. The yellow line is the path of the second shock front. The red solid, red dashed, and green dashed line are the merged shock front, the contact discontinuity, and the adiabatic release, respectively.
\textbf{(b).} Schematic picture of the shocked target for $t_1 < t < t_2$ (b1), $t_2 < t < t_3$ (b2), and $t_3 < t < t_4$ (b3). \textit{(b1).} The first shock propagates at $U_s^0$, leaving a transparent shocked state 1 moving at $U_p^1$.  \textit{(b2).} The second shock propagates at $U_s^1$, leaving a double-shocked state 2 moving at $U_p^2$. \textit{(b3).} The merged shock propagates at $U_s^m$, leaving a shocked material in state 3o. A cold front divides 3o and 3i. State 2 still exists behind and a gradual adiabatic transition connects the regions 2 and 3i. The dashed dark-yellow circles indicate the quantities measured during each time window.
}
\label{fig:scheme}
\end{figure*}

\section{Principles of the method}
\label{sec:technique_principles}

Our technique is based on the sequential propagation of two shocks inside the considered sample. A similar method has been demonstrated in a previous experiment \cite{benuzzi04}. In that case, the shocks were propagating inside a metallic sample so that the VISARs could only measure its free-surface velocity. The reached thermodynamic state was inferred using an equation of state model of the shocked material. In our case, the characterisation of the sample is more direct since it is model-independent and based on shock and material velocity measurements.

In our approach, the first shock brings the material to a uniformly compressed denser state. To allow precise characterisation of the achieved thermodynamic conditions, we require the material to be kept in a transparent state. The extension to opaque regimes is still possible, but it demands a different analysis and will not be presented here. In order to satisfy such requirements, the first shock must be sustained by a long (10 ns) and weak (intensity of $\sim 10^{12}$ W/cm$^2$) laser pulse. The second shock aims at creating the high pressures conditions and is sustained by a much more intense ($\sim 10^{13}$ W/cm$^2$), but shorter (2 ns) laser pulse. Since it loads a material with a higher density than the ambient one, it generates colder states with respect to those belonging to the principal Hugoniot. Given that the second shock is faster than the first, the two eventually merge inside the target. The merged shock brings back the sample to a state on the principal Hugoniot.

The thermodynamic state of the double-shocked sample is measured via a ``self impedance mismatch'' technique. In practice, this requires the knowledge of the first-shocked and merged-shocked states (both lying along the principal Hugoniot) and of the second-shock velocity. 

\subsection{Hydrodynamic path}

The hydrodynamic path of the shocks inside the sample is shown in Figure \ref{fig:scheme}(a). 

The first shock enters the sample at $t = t_1$, propagating at the velocity $U_s^0 (t)$. The shocked state 1, which moves at the material velocity $U_p^1 (t)$, is optically transparent as we required. Therefore, we cannot directly measure $U_s^0 (t)$ with VISARs. The probe laser is instead reflected by the interface between the sample and a metal layer designedly placed before it. The interface is moving at the material velocity of the shocked state 1 ($U_p^1 (t)$) since the sample and the interface itself are at mechanical equilibrium. Therefore at $t = t_1$ the VISAR diagnostics detect a fringe shift allowing the measurement of $U_p^1 (t)$. If the first laser pulse is properly designed to guarantee stationarity of the first shock, the material velocity is constant ($U_p^1 (t) \simeq U_p^1$) and the VISAR provides with $U_p^1$ (see Section \ref{sec:up1}).

At $t = t_2$, the second shock enters the sample moving at $U_s^1$. The sample is brought to the unknown double-shocked state 2. A second fringe shift or a fringe extinction is noticeable on VISARs, depending on the optical properties of state 2. Since the second shock is faster than the first, the two will merge at $t = t_3$. When this happens, a discontinuity between the un-shocked state 0 and the double-shocked state 2 occurs. This transition is not consistent with the Rankine-Hugoniot relations. Indeed, the internal energy of the double-shocked state 2 is lower than that of a state at the same pressure but obtained through a single-shock compression. Therefore, an adiabatic release takes place from the double-shocked region 2 to a newly formed region 3. The region 3 divides into two sub-regions 3i and 3o, separated by a contact discontinuity, across which density and internal energy are not continuous, while pressure and fluid velocity are. The sub-region 3i is the result of the adiabatic release from state 2, while 3o is consistent with the Rankine-Hugoniot relations from state 0 and therefore lies along the principal Hugoniot \cite{birnboim10}.

At $t = t_3$, a strong emission increase is observed by the SOP. Indeed, the state behind the shock front lies along the principal Hugoniot. The pressure behind the merged shock is slightly lower than that of the second shock, as a consequence of the adiabatic release, but temperature is much higher and can typically be detected.
Moreover, since high pressures are aimed for the state 2, at the only slightly lower pressure on the Hugoniot the shock front is typically optically reflecting and shock velocity measurements can also be achieved with the VISARs. The merged-shock front velocity decreases in time because the shock is generally no longer sustained by the laser pulse. As a consequence, also the emission is fading in time up to $t_4$, when the merged shock exits the sample and the signal suddenly drops. 

Schematic pictures of the target structure for $t_1 < t < t_2$, $t_2 < t < t_3$, and $t_3 < t < t_4$ are shown in Figure \ref{fig:scheme}(b). 

\subsection{Self impedance mismatch}
\label{sec:SIM}

\begin{figure}[h!]
\includegraphics[width=0.9\columnwidth]{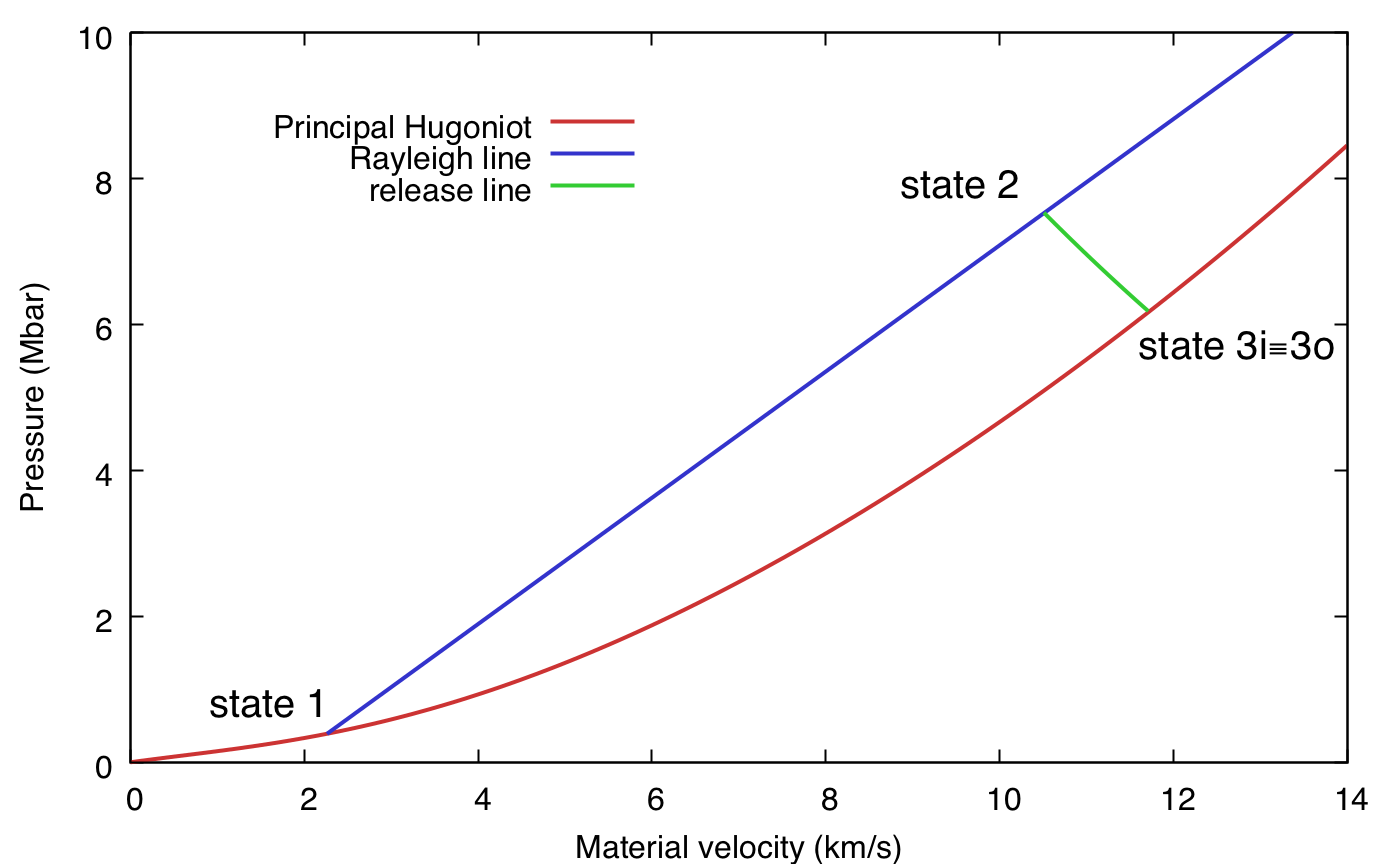}
\caption{Self impedance mismatch analysis in the $(U_p, P)$ plane for the determination of the double-shocked state. State 3 corresponds to both states 3i and 3o, which share the same pressure and material velocity.}
\label{fig:doubleshock_IMA}
\end{figure}

The impedance mismatch technique \cite{forbes} is a common procedure in the field of dynamic compression for measuring the thermodynamic state of an unknown material. A standard material is placed before the sample and the shock velocity is measured in the two materials. This technique requires the knowledge of the equation of state of the standard material and is based on conservation laws at the shocked material interface to infer the thermodynamic state reached in the sample. In our framework, the state of the double-shocked region 2 is obtained by applying the impedance mismatch technique to the different regions of the $\alpha$-quartz sample, with the well characterised region 1 and 3o used as standard. Since impedance mismatch is not applied to different materials, but to different states of the same sample, this procedure is denoted hereafter as ``self impedance mismatch''.

Figure \ref{fig:doubleshock_IMA} illustrates the procedure to determine the double-shocked state in the $(U_p, P)$ plane. The principal Hugoniot of the sample is traced and the state 1, which lies on it since it is obtained with compression by a single shock, is individuated by $U_p^1$, measured through the VISARs (see Section \ref{sec:up1}).

A Rayleigh line 
\begin{equation}
P(U_p) = P_1 + \rho_1 \left(U_s^1 - U_p^1 \right) \left( U_p - U_p^1 \right)
\end{equation}
(where $P_1$ and $\rho_1$ are the pressure and density of the state 1, respectively) is then traced starting from state 1. This line expresses the momentum conservation along the second shock front. The state $\left(U_p^2, P_2 \right)$ reached behind the second shock must thus belong to this line.

State 3o, reached behind the merged shock right after its formation, is determined by the measure of $U_s^m (t_3)$ (see Section \ref{sec:measure_usm}). Since state 3o lies along the principal Hugoniot, knowing its shock velocity allows to determine $(U_p^{3o}, P_{3o})$. State 3o has the same pressure and material velocity of state 3i, which is the result of an adiabatic release starting from state 2. To model the transition from state 2 to state 3i, an adiabatic compression path must be calculated starting from 3o (the Hugoniot state corresponding to 3i in pressure and material velocity). A Mie-Gr\"uneisen model has been established for $\alpha$-quartz \cite{knudson13} in the $3 - 12$ Mbar range. The intersection between the adiabatic path and the Rayleigh line gives the equation-of-state point in the $(U_p, P)$ plane for the double-shocked state 2. Density $\rho_2$ is then determined through the generalised Rankine-Hugoniot equation stating mass conservation:
\begin{equation}
\label{eqn:rho_2}
\rho_2 = \rho_1 \frac{U_s^1 - U_p^1}{U_s^1 - U_p^2}.
\end{equation}
Finally, the internal energy density $E_2$ is obtained using the Rankine-Hugoniot equation for energy conservation:
\begin{equation}
\label{eqn:E_2}
E_2 = E_0 + \frac{P_2 U_p^2 - P_1 U_p^1}{\rho_1 \left( U_s^1 - U_p^1 \right)} + \frac{1}{2} \left[ \left( U_p^1 \right)^2 - \left( U_p^2 \right)^2 \right],
\end{equation}
where $E_0$ is the initial internal energy density.

\begin{figure}[h!]
\includegraphics[width=0.9\columnwidth]{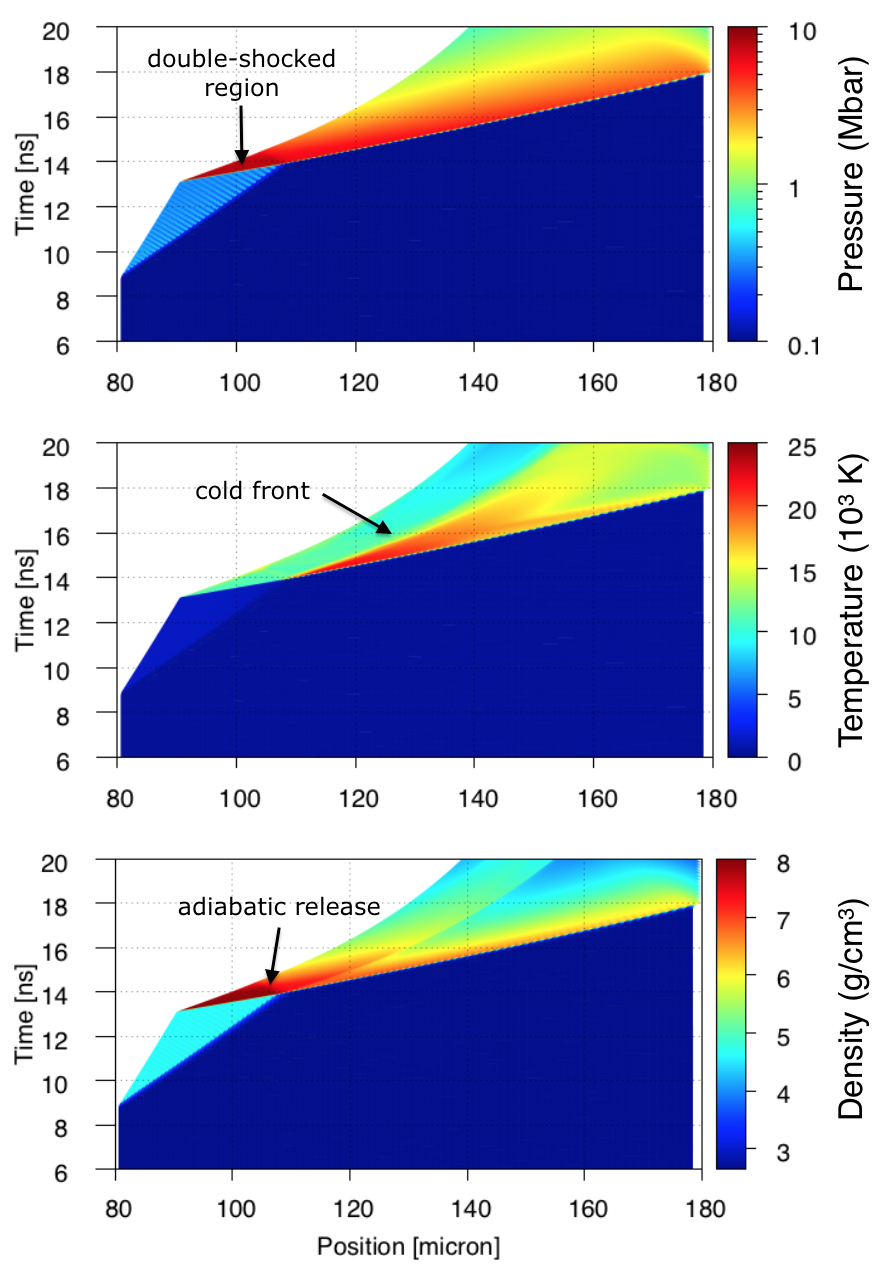}
\caption{Pressure, temperature, and density (in color scale) history inside the $\alpha$-quartz sample. The drive laser comes from the left, diagnostics are on the right side.}
\label{fig:multi_x}
\end{figure}

\begin{figure}[h!]
\includegraphics[width=0.9\columnwidth]{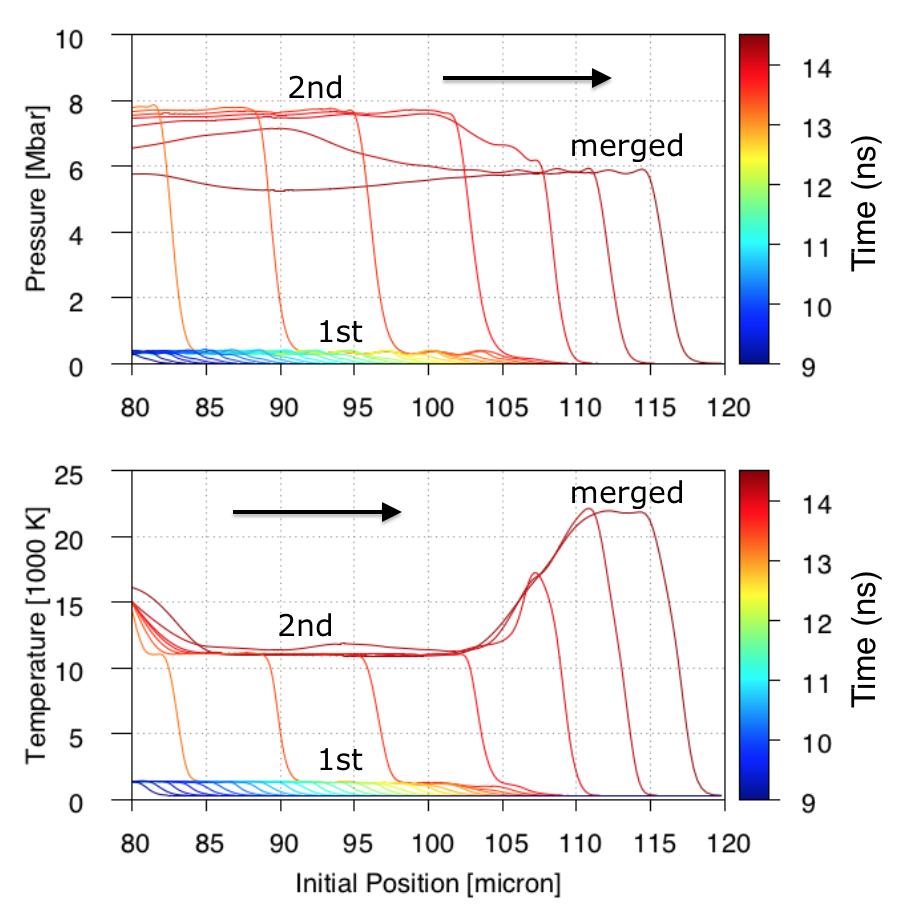}
\caption{Pressure and temperature profile in a portion of the $\alpha$-quartz sample depending on time (color scale). The drive laser comes from the left, diagnostics are on the right side. The arrows indicate the shock propagation direction. The loading pressure behind the merged shock is slightly lower than behind the second shock. On the contrary, temperature of portion loaded by the merged shock, which lies along the principal Hugoniot, is sensibly higher than the second-shock loaded region, which is an off-Hugoniot state.}
\label{fig:multi_t}
\end{figure}

\section{Simulations and design of the experiment}
\label{sec:simulations}

The experimental application of the double-shock compression technique requires the fulfilment of several conditions: \textit{(i)} the pressure reached with the first shock loading must be low enough for the sample to remain transparent; \textit{(ii)} the second shock lifetime must be long enough to allow optical probing, but short enough for considering it as constant in time; \textit{(iii)} the merging of the two shocks must take place before the exit of the first shock from the sample; \textit{(iv)} no unwanted shock reverberations must occur, which would affect the state reached by shock loading. Moreover, a rough estimation of the timings of the process, from the entering of the first shock to the exit of the merged shock, must be made to set the diagnostics delays before the shot. The reached thermodynamical conditions must also be previously estimated.

Hydrodynamical simulations have been performed with the Lagrangian one-dimensional code \textit{MULTI} \cite{ramis88} to optimise the laser pulse shape and the target structure in order to meet the above requirements.

The target employed in the simulations consisted in a plastic layer as ablator, an aluminum layer, and the $\alpha$-quartz sample. The aluminum layer allows both transmitting a controlled shock to the quartz sample and measuring its material velocity when in state 1 by reflecting the probe laser at the aluminum/quartz interface. The equations of state employed for the simulations has been taken from \textit{SESAME} \cite{lyon92, johnson94}. Table 7592, 33717, and 7360 were used for plastic, aluminum, and quartz, respectively.

Figures \ref{fig:multi_x} and \ref{fig:multi_t} show the output of a simulation using a target composed by 10 $\mu$m of parylene, 70 $\mu$m of aluminum, and 100 $\mu$m of $\alpha$-quartz and two laser pulses, the first starting at 0 ns with an intensity of $0.135 \times 10^{13}$ W/cm$^2$ and a duration of 10 ns, the second starting at 10 ns with an intensity of $4.2 \times 10^{13}$ W/cm$^2$ and a duration of 2 ns.

The thickness of the aluminum layer has been chosen to avoid shock reverberations, as imposed by requirement \textit{(iv)}. The first laser pulse duration is long enough to meet criterion \textit{(ii)}, and its low intensity allows to respect the requirement \textit{(i)}. The high intensity of the second pulse let us comply with criteria \textit{(ii)} and \textit{(iii)}. Furthermore, simulations allow to anticipate the timings and the thermodynamical conditions reached in the sample.

Figure \ref{fig:multi_x} shows the simulated time-dependent pressure, temperature, and density (in color scale) profiles inside an $\alpha$-quartz sample. The $x$ axis is the position in the laboratory frame. The white region on the left of the sample is occupied by aluminum, which has been excluded by the plots for clarity. The action of the first shock can be noticed from 9 ns. The aluminum/quartz interface also starts moving. The second shock enters quartz at around 13 ns, engendering a sudden change in the interface velocity. Between 9 and 14 ns, the pressure and temperature profiles present two successive jumps. The two shocks merge at around 14 ns. An adiabatic release then follows from the double-shocked region to the region behind the merged shock. The release is localised at around 110 $\mu$m and can clearly be seen in the pressure and the density plot. A contact discontinuity is generated and propagates at the material velocity. This feature is observed as a discontinuity in the temperature and density plots, but it does not exist in the pressure profiles since pressure and material velocity are conserved.

Figure \ref{fig:multi_t} shows the pressure and temperature evolution (time in color scale) in the first 40 $\mu$m of the sample. The $x$ axis is the initial position (\textit{i.e.} the Lagrangian variable). 

In this representation we find the same salient stages as explained above. The first weak shock enters the sample at 9 ns and generates low pressure - low temperature uniform conditions. It is followed by the second shock at 14 ns. The  space profile of pressure and temperature starting from this time consists in two successive jumps, with the highest discontinuity pursuing the first slower one. Their merging at 14 ns is principally associated to a huge increase in temperature and a slight decrease in pressure. The uniform pressure and temperature conditions depicted in Figure \ref{fig:multi_t} show the excellent stationarity of both shocks.

We demonstrated the feasibility of the method by performing some shots at the LULI2000 laser facility, using the target and laser pulse specifications determined in Section \ref{sec:simulations}.

\begin{figure*}
\includegraphics[width=0.8\textwidth]{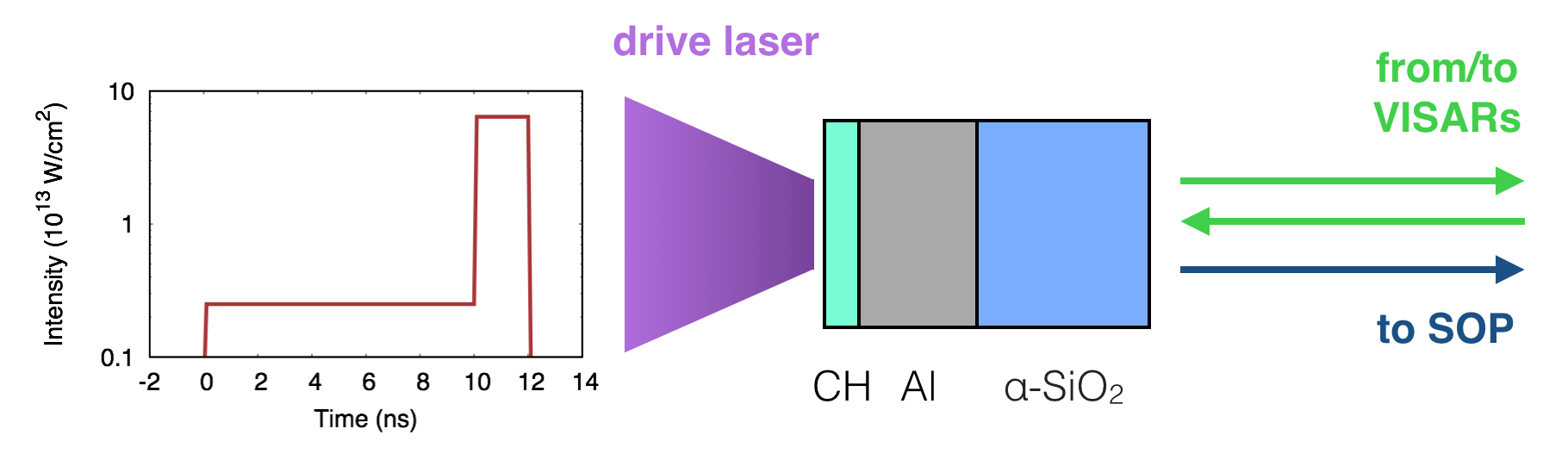}
\caption{Time profile of the drive-laser intensity, focusing of drive laser onto the target, target structure scheme, and simplified scheme of the optical path from and to the diagnostics.}
\label{fig:laser_target_diag}
\end{figure*}

\section{Experimental application of the method}
\label{sec:exp_application}

\subsection{Setup}

Two distinct laser pulses at 527 nm, obtained with the LULI2000 North and South chains, have been used to generate the two shocks inside the sample. The first pulse was $10$ ns long and of weak, constant intensity ($\sim 10^{12}$ W/cm$^2$). The second pulse, delivered right after the end of the first one, was $2$ ns long and much more intense ($\sim 10^{13}$ W/cm$^2$). Phase plates were used to irradiate the target with uniform intensity on a focal diameter of 500 $\mu$m. Figure \ref{fig:laser_target_diag} shows the time profile of the drive-laser intensity and the focusing of the drive laser onto the target.

Targets complied with the design optimised in the hydrodynamic simulations discussed above. They were made by a 10 $\mu$m CH, 70 $\mu$m Al pusher, glued on a 100 $\mu$m thick z-cut $\alpha$-quartz sample (see Figure \ref{fig:laser_target_diag}).

Optical diagnostics typical of shock experiments were set up. They included two VISAR interferometers working at 532 nm and 1064 nm and used to measure shock or material velocities. VISAR sensitivities were set to 15.94 km/s (at 1064 nm) and 6.11 or 3.31 km/s (at 532 nm). The probe laser pulse duration were 20 ns and 10 ns respectively. Streak optical pyrometer (SOP) was also implemented in the set-up. It collected self-emission from the target on a 20 nm wide spectral window centered around $\lambda_0 = 455$ nm. Typical temporal windows for both diagnostics were 10 or 20 ns.

\begin{figure}[h!]
\includegraphics[width=0.9\columnwidth]{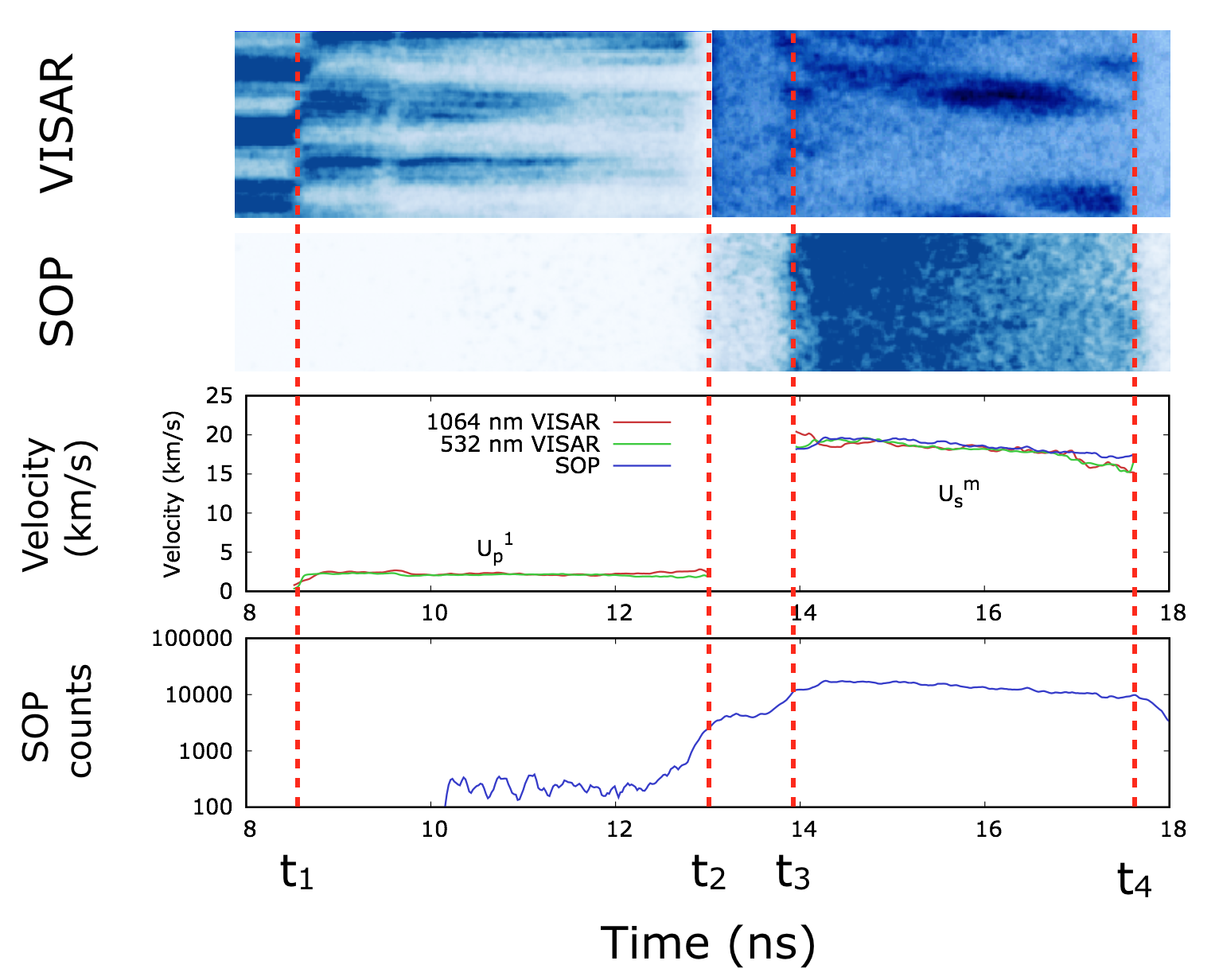}
\caption{Output image of the 532 nm VISAR (first item) and of the SOP (second item) for a typical shot. $U_p^1 (t)$, measured with the two VISARs, and $U_s^m (t)$, measured with the VISARs and with the SOP signal converted through the known $\alpha$-quartz $T(U_s)$ relation \cite{millot15}, are shown in the third item. Time-dependent SOP counts are shown in the fourth item. The color scale of the VISAR image has been changed for $t > t_2$ to increase readability.}
\label{fig:diag_output}
\end{figure}

\subsection{Analysis of diagnostics outputs}
\label{sec:diagnostics_outputs}

VISARs and SOP output images (see Figure \ref{fig:diag_output}) have been analysed using the \textit{Neutrino} software \cite{flacco11} to obtain time-dependent fringe shift, intensity, and self-emission. Here we describe the analysis we performed to extract information about velocities, temperatures, and second shock front reflectivity.

\subsubsection{Material velocity after the first shock}
\label{sec:up1}

Particle (material) velocity of the first shocked state was directly inferred by the fringe shift $\Delta \phi (t)$ of the VISARs in the $t_1 - t_2$ time interval:
\begin{equation}
U_p^1 (t) = \frac{\textrm{VPF}}{n^\star} \frac{\Delta \phi (t)}{2 \pi},
\end{equation}
where $\textrm{VPF}$ is the velocity-per-fringe parameter of the VISAR and
\begin{equation}
n^{\star} = n_0 - \rho_0 \left( \frac{dn}{d\rho} \right)_{\rho = \rho_0}
\end{equation}
(where $n_0$ and $\rho_0$ are the initial refractive index at the probe laser wavelength and the initial density, respectively) is a correction which takes into account the refractive index change on the shocked part of the sample, assuming a linear dependence on density of the refractive index \cite{hardesty76, setchell79}.

In the experiment shot, we obtained a mean material velocity $U_p^1 = 2.20 \pm 0.11$ km/s, corresponding to a pressure $P_1 = 0.33 \pm 0.02$ Mbar.

\subsubsection{Merged-shock velocity}
\label{sec:measure_usm}

The time-dependent merged-shock velocity is measured from the VISAR fringe shift as in the classical case of strong reflecting shocks:
\begin{equation}
U_s^m (t) = \frac{\textrm{VPF}}{n_0} \frac{\Delta \phi (t)}{2 \pi},
\end{equation}
where $n_0$ is the $\alpha$-quartz initial refractive index at the probe laser wavelength. 

In practice, the interaction between the two shock fronts may affect the flatness of the resulting shock front. As a result the quality of the merged shock VISAR signal may decrease. To overcome this limitation, the SOP signal can be used to indirectly determine $U_s^m (t)$ if the $T - U_s$ relation along the principal Hugoniot is known. 

In our experiment shot, we found $U_s^m (t_3) = 19.9 \pm 1.0$ km/s from a fit on the VISAR signal and $U_s^m (t_3) = 20.4 \pm 1.1$ km/s from a fit on the merged shock velocity indirectly obtained from the SOP signal. These values are well compatible within errors.

The mean merged-shock velocity
\begin{equation}
\label{eqn:usm_mean}
\left< U_s^m \right> = \frac{1}{t_4 - t_3} \int_{t_3}^{t_4} U_s^m (t) dt = \frac{x_4 - x_3}{t_4 - t_3}
\end{equation}
is measured from the VISAR and SOP timings $t_3$ and $t_4$ (see Figure \ref{fig:diag_output}), the calculated $x_3$ and the initial position of the rear surface of the sample $x_4$. This quantity is critical for SOP calibration (see Section \ref{sec:measure_temp}).

\subsubsection{Second-shock velocity}

In the case the second shock is reflecting, its shock velocity is directly deduced from the VISAR. In the other cases, including the situation where the fringes are not of sufficient quality to allow fringe-shift analysis, the $U_s^1$ is inferred from  measure $U_p^1$ and $U_s^m (t_3)$ measurements and from the timings $t_1$, $t_2$, and $t_3$.

Let $x_1$ be the initial position of the aluminum/quartz interface in the laboratory frame, $x_2$ its position when the second shock enters the sample, and $x_3$ the position of the two shock fronts when they merge.
It follows that
\begin{equation} 
x_2 - x_1 = U_p^1 (t_2 - t_1) 
\end{equation}
\begin{equation}
x_3 - x_1 = U_s^0 (t_3 - t_1),
\end{equation}
where $U_{s}^0$ (the first shock velocity) is given by the known $U_s - U_p$ relation in quartz.
The second shock velocity $U_{s}^1$ can be found as follows:
\begin{equation}
U_{s}^1 = \frac{x_3 - x_2}{t_3 - t_2}.
\end{equation}
Assuming that the second shock velocity is constant, which is reasonable because of the short lifetime of the second shock (1 - 2 ns, as corroborated by simulations). We determined $U_s^1 = 22.3 \pm 2.8$ km/s for our test shot.

\subsubsection{Second-shock-front reflectivity}
\label{sec:measure_R}

Shock loading of a transparent material changes the real part of its refractive index $n$ due to the density increase. Moreover, if the shock compression is strong enough, the loaded sample becomes conducting \cite{clerouin05, french11}, thus the imaginary part of its refractive index $k$ cannot be neglected anymore. The global refractive index change makes the shock front reflecting according to the Fresnel's law.

In our case, the second-shock front propagates through state 1 and its reflectivity will be
\begin{equation}
R = \left| \frac{\tilde{n}_2 - \tilde{n}_1}{\tilde{n}_2 + \tilde{n}_1} \right|^2,
\end{equation}
where $\tilde{n}_1$ and $\tilde{n}_2$ are the complex refractive indices of the region loaded by the first and second shock, respectively.

If the refractive index behaviour of the sample is known at the conditions reached in state 1, so is $\tilde{n}_1$ and hence the measure of $R$ can be used to estimate $\tilde{n}_2$. The electrical conductivity can then be estimated, \textit{e.g.} via a Drude-Sommerfeld model \cite{celliers00, celliers10, millot18}.

If the probe laser is reflected by a shock front propagating into a transparent sample, shock-front reflectivity at the probe laser wavelength can be obtained as the ratio of the VISAR signal acquired during the shot to a reference one. In the framework of double-shock compression, the second and the merged shock front reflectivity can be measured. This procedure can fail in giving an absolute reflectivity measurement if the reference metallic layer is oxidised or contaminated in the gluing process.

If the sample reflectivity is well characterised as a function of the shock velocity on the Hugoniot curve, as it is for quartz \cite{hicks06, millot15}, the measured reflectivity of the merged shock front $R(U_s^m (t))$ can be used as a calibration factor. 

\subsubsection{Temperatures}
\label{sec:measure_temp}

If the shock front is propagating in an ideally transparent material, \textit{i.e.} without being partially absorbed, the temperature of the shocked sample can be evaluated from the SOP counts through an inverse Planck's law. The emissivity at the SOP wavelength is estimated from the reflectivity at the probe wavelength measured by the VISARs (see Section \ref{sec:measure_R}) via a grey-body hypothesis \cite{bolis16}.

The SOP is calibrated before the experiment, using a lamp of known emission temperature. Shot-by-shot \textit{in situ} calibration can also be obtained using the known temperature - shock velocity relation of a standard placed inside the target and the shock velocity measurements from the VISARs. The second method is possible in the framework of double-shock experiments if the $T(U_s)$ relation of the sample along the principal Hugoniot is known. In this case, in the $t_3 - t_4$ time interval, $U_s^m (t)$ is measured from VISARs and a SOP calibration is obtained applying the $T(U_s^m (t))$ relation.  
The known $T (U_s)$ relation can also be inversely used to obtain the initial merged shock velocity $U_s^m (t_3)$, which is needed for self impedance mismatch (see Section \ref{sec:SIM}), from the temperature measurement.

The temperature measurement of the double-shocked region must be carefully handled. Indeed, the second shock front propagates in a sample which has been pre-compressed by the first shock and may thus be partially absorbing. Defining $\mu_1$ as the absorption coefficient of the sample loaded by the first shock, the time-dependent intensity collected by the SOP (proportional to the number of counts) $I_{collected} (t)$ will be
\begin{equation}
\label{eqn:int}
I_{collected}(t) = I_{emitted}(t) \exp \left[ - \mu_1 l_1 (t) \right],
\end{equation}
where $I_{emitted}(t)$ is the time-dependent intensity emitted by the double-shocked sample and $l_1 (t)$ is the width of the sample region already loaded by the first shock but not loaded by the second one yet. 

Geometrical considerations in the propagation direction - time plane (see Figure \ref{fig:scheme}(a)) involve that:
\begin{equation}
\label{eqn:length}
l_1 (t) = U_s^0 (t - t_1) - U_s^1 (t - t_2) - (x_2 - x_1).
\end{equation}
An exponential fit on the SOP counts using equations (\ref{eqn:int}) and (\ref{eqn:length}) is performed to measure the emitted intensity and thus the temperature of the double-shocked sample. In principle, if the time resolution of the SOP is high enough, the fit will also provide a reliable measure of the absorption coefficient $\mu_1$. For our test shot, we measured $T_2 = \left(16.5 \pm 3.2 \right) \times 10^3$ K, but we could not estimate $\mu_1$ as the SOP time resolution was not sufficient.

\begin{figure}[h!]
\includegraphics[width=0.9\columnwidth]{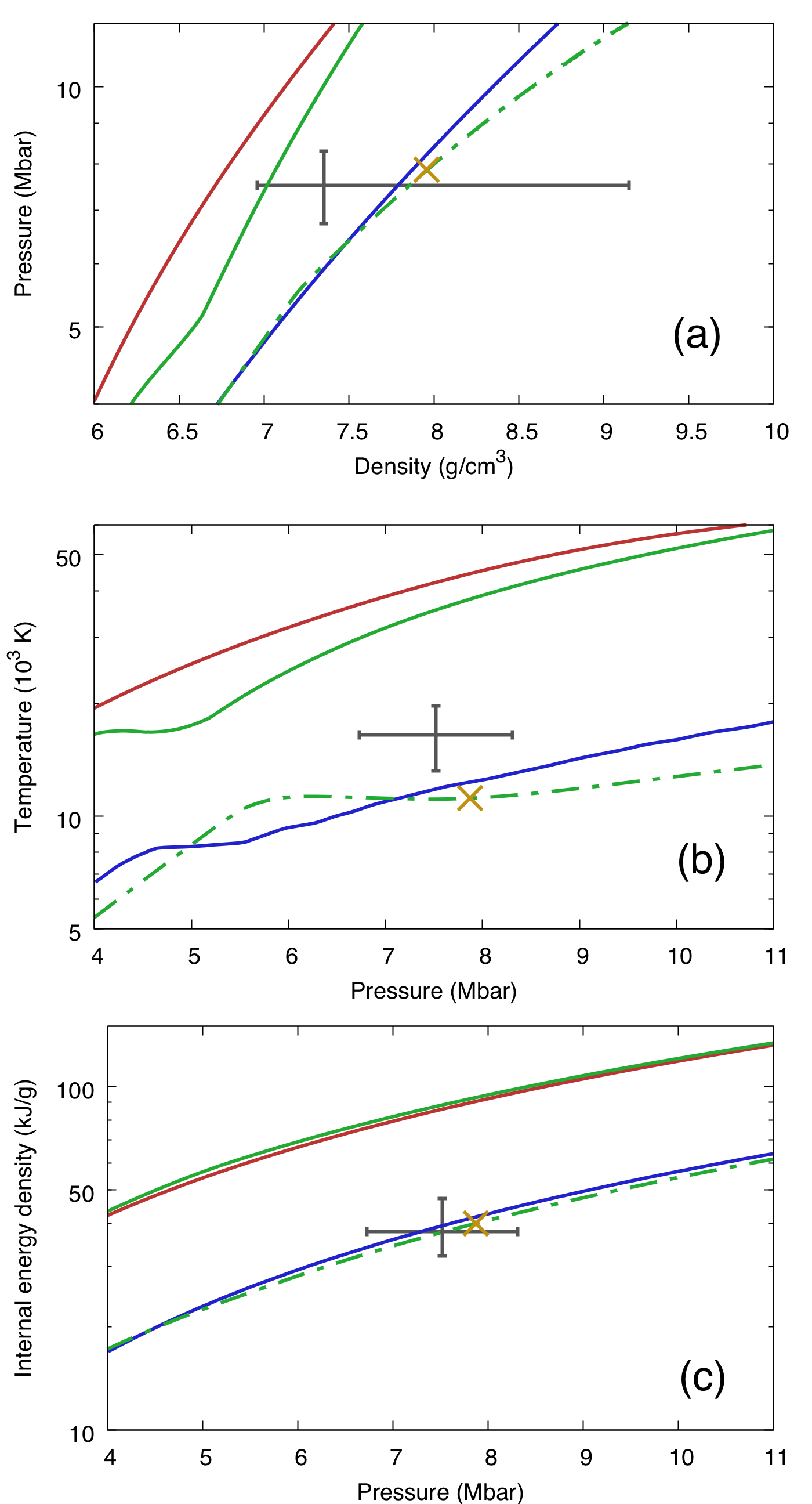}
\caption{Pressure - density (a), temperature - pressure (b), and internal energy density - pressure (c) conditions reached in the test shot (grey datum with errorbars). Experimentally determined $\alpha$-quartz and stishovite principal Hugoniot lines (red and blue line, respectively \cite{millot15}) and principal and pre-compressed Hugoniot lines of $\alpha$-quartz from \textit{SESAME} table 7360 (solid and dashed-dotted green line, respectively) are shown for comparison. Conditions reached via the \textit{MULTI} simulation are also shown (yellow cross).
}
\label{fig:res_multiplot}
\end{figure}

\section{Results and discussion}
\label{sec:results}

The output of a VISAR image and of the SOP image for a test shot on $\alpha$-quartz is shown in Figure \ref{fig:diag_output}. The time profile of the first material velocity and of the merged shock velocity ($U_p^1 (t)$ and $U_s^m (t)$, respectively) and of the SOP counts are also shown. The red dotted lines correspond to the timings $t_1$, $t_2$, $t_3$, and $t_4$. In the $t_1 - t_2$ time window, VISAR fringes are shifted with respect to the unperturbed position and are fairly constant in time, while the SOP signal is very low ($\sim 100$ counts) and noisy. From $t_2$ to $t_3$, the VISAR fringe system disappears leaving a single shifted fringe surrounded by a noisy region. The SOP signal gradually increases around $t_2$ and stabilises for the duration of the second shock. At $t = t_3$, a fringe system reappears on the VISAR image. Fringes move in time in the $t_3 - t_4$ time window, since the merged shock is unsustained by the laser pulse and thus decaying in time. Around $t = t_3$ an important increase is noticeable in the SOP counts, due to the fact that the emitting sample now lies along the principal Hugoniot. The SOP counts then decay in time since the shock is losing its energy. The color scale of the VISAR image in Figure \ref{fig:diag_output} has been changed for $t > t_2$ to increase readability. Indeed, fringes in the $t_1 - t_2$ window are the result of a reflection on a metallic surface, while from $t_2$ to $t_4$ their intensity is much lower since they are the result of a shock-front reflection.

The spatial structure of VISAR fringes is non-uniform starting from $t_2$. We attribute this to spatial modulations due to the interaction between the two shocks, that amplify previous non-planarities in the spatial profile of each shock. We remind that it is important to say that possible issues in measuring the merged-shock velocity due to fringe disappearing would not be critical for the analysis, since $U_s^m (t)$ can be indirectly measured from the SOP signal (see Section \ref{sec:measure_temp}).

Figure \ref{fig:res_multiplot} places the double-shocked state obtained from the shot of Figure \ref{fig:diag_output} in the phase space (in the $P - \rho$ (a), $T - P$ (b), and $E - P$ (c) planes) and compares it with the principal Hugoniot lines of $\alpha$-quartz and stishovite \cite{millot15}. The pressure reached after the first shock loading was $P_1 = 0.33 \pm 0.02$ Mbar. The temperature of the double-shocked state was $T_2 = \left(16.5 \pm 3.2 \right) \times 10^3$ K at $P_2 = 7.52 \pm 0.79$ Mbar, $60\%$ lower than the one along the principal Hugoniot of $\alpha$-quartz at the same pressure, and higher than along the principal Hugoniot of stishovite. The density of the double-shocked state was $\rho_2 = 7.35^{+ 1.80}_{- 0.39}$ g/cm$^3$ and the internal energy density jump $E_2 - E_0 = 37.8^{+ 9.5}_{- 5.6}$ kJ/g. 

Uncertainties on $\rho$, $P$, $E$, and $T$ have been estimated through a Monte-Carlo routine. The high uncertainty on density is due to the small value of the denominator in equation \ref{eqn:rho_2}. This uncertainty is highly asymmetric because of the requirement that the density of a double-shocked state must be higher than the Hugoniot one at the same pressure. The measured and simulated temperature of state 2 are not compatible within errors. This is due to the fact that simulations used an equation of state table for quartz that does not reproduce temperatures correctly, as it can be seen comparing the Hugoniot lines in Figure \ref{fig:res_multiplot}(b). The simulated temperature is lower than the actual one along both the principal and pre-compressed Hugoniot. On the contrary, the table used in the simulations reproduces quite well the experimental internal energy density (Figure \ref{fig:res_multiplot}(c)), and even for state 2 the measured and simulated values are compatible within errors.

As shown in Figure \ref{fig:res_multiplot}, the double-shock technique allowed us to reach thermodynamic conditions between the Hugoniot lines of $\alpha$-quartz and stishovite. By changing the timings and the relative intensities between the two laser pulses, a large part of the aforementioned phase space region could be mapped. 

\section{Conclusions}
\label{sec:conclusions}

We applied a double-shock technique in a well characterised framework that allows reaching and probing multimegabar states at lower temperatures than along the principal Hugoniot. We experimentally demonstrated its feasibility on $\alpha$-quartz.

In future works, this technique should be extensively applied silica and to other components of planetary interiors to explore low-temperature phases and phase boundaries relevant for planetology. Moreover, it should be generalised to the case of three or more shocks to obtain even colder states, thus exploring a wider range of thermodynamical conditions.

\begin{acknowledgments}

We want to thank the LULI support staff and the LULI2000 laser facility staff for their invaluable contribution to the experiments. We are grateful to Tom Boehly for the useful discussions. This research has been supported by the French \textit{Agence Nationale de la Recherche} funding ANR-16-CE31-0008 (POMPEI).

\end{acknowledgments}

\end{document}